\documentclass[10pt,twocolumn]{article}
\usepackage{graphicx}
\usepackage{color}
\begin{document}

\title{Quantum temporal imaging: application of a time lens to quantum optics}

\author{G. Patera$^1$, J. Shi$^{1,2,3}$, D. B. Horoshko$^{1,4}$, M. I. Kolobov$^1$\vspace{10pt}\\
$^1$Univ. Lille, CNRS, UMR 8523 - PhLAM - Physique des Lasers Atomes et Mol\'{e}cules, \\F-59000 Lille, France\\
$^2$Key Laboratory for Quantum Optics, Shanghai Institute of Optics and Fine Mechanics, \\Chinese Academy of Sciences, Shanghai 201800, China\\
$^3$University of Chinese Academy of Sciences, Beijing 100049, China\\
$^4$B.~I.~Stepanov Institute of Physics, NASB, Nezavisimosti Ave.~68, \\Minsk 220072 Belarus}

\maketitle

\begin{abstract}
We consider application of a temporal imaging system, based on the sum-frequency generation, to a nonclassical, in particular, squeezed optical temporal waveform. We analyze the restrictions on the pump and the phase matching condition in the summing crystal, necessary for preserving the quantum features of the initial waveform. We show that modification of the notion of the field of view in the quantum case is necessary, and that the quantum field of view is much narrower than the classical one for the same temporal imaging system. These results are important for temporal stretching and compressing of squeezed fields, used in quantum-enhanced metrology and quantum communications.
\end{abstract}

\section{Introduction}

Optical temporal imaging is a technique of stretching or compressing an optical waveform in time, while fully preserving its temporal structure \cite{Copmany2011,Salem2013}. This technique shares much in common with the optical spatial imaging, aiming at stretching or compressing the spatial distribution of the optical wave in a given area \cite{Kolner1994}. In the last decades the optical temporal imaging became an elaborated technique with various applications, mainly dealing with temporal stretching of ultrafast waveforms with bandwidth of tens and hundreds THz by several orders of magnitude, in order that these waveforms are detectable by ordinary photodetection systems with a bandwidth not surpassing 30 GHz. Another important application is compression of waveforms created by electro-optical devices to picosecond or sub-picosecond timescales, for their transmission with a highly increased data rate. The key element of a temporal imaging system is time lens, introducing a quadratic in time phase modulation into the signal field. Optical time lenses presently are based on electro-optical phase modulation \cite{Giordmaine1968,Grischowsky1974,Kolner1988,Kolner1989,Karpinski2016}, sum-frequency generation (SFG) \cite{Agrawal1989,Bennett1994,Bennett1999,Bennett2000a,Bennett2000b,Bennett2001,Hernandez2013}, or four-wave mixing \cite{Foster2008,Foster2009,Okawachi2009,Kuzucu2009} and provide a temporal magnification up to 100 times.

The conventional theory of optical temporal imaging treats the electromagnetic field as a classical signal. However, the quantum nature of light becomes important in some special cases, in particular when the information on some physical process is conveyed by single photons or by squeezed light. The photodetection techniques utilized in these cases -- single-photon detection and homodyne detection respectively -- share the same bandwidth limitation of about 30 GHz. As in the classical optics, the application of temporal imaging to the signals carried by a quantum light would allow one to observe directly these signals on a much shorter timescale. However, the corresponding theory should be explicitly based on the quantum formalism. Such generalisation of classical temporal imaging into quantum regime is natural to call quantum temporal imaging. It is appropriate to mention here the conventional (spatial) quantum imaging \cite{Lugiato2002,QI-JMO2006,Shih2007,Kolobov2007}, which has become a well-developed branch of quantum optics. Quantum temporal imaging can borrow many ideas and techniques from its spatial counterpart. Several approaches to the quantum temporal imaging at the single-photon level have been published in the last decade \cite{Kielpinski2011,Lavoie2013,Zhu2013,Donohue2015}, but the case of continuous variables has been considered only recently \cite{Patera2015}. It has been shown that having a squeezed light as the object imposes additional constraints on the temporal imaging system, if one aims at preserving the quantum structure of the object, namely, its degree of squeezing.

The aim of the present article is to develop the approach of Ref.~\cite{Patera2015} and to analyze in detail the constraints imposed on the temporal imaging system in a quantum regime. As in Ref.~\cite{Patera2015}, here we consider a temporal imaging system based on the SFG in a nonlinear optical crystal, having the signal beam in a squeezed state as one input and a strong coherent beam, a pump, as another input.

The article is organized as follows. In Sec.~2 we briefly review the principles of quantum temporal imaging with squeezed light and consider the effect of temporal magnification and non-unit efficiency of the imaging system on the spectrum of squeezing. In Sec.~3 we  analyze the restrictions imposed on the pump and the phase matching of the waves in the crystal in the quantum regime and discuss the modification of the notion of the field of view in the quantum case. The main results of the article are summarized in Sec.~4.

\section{Time lens based on the sum-frequency generation}

\subsection{Field transformation}

{
In this section we consider a time lens based on the SFG process in a $\chi^{(2)}$ nonlinear medium~\cite{Bennett2000a,Bennett2000b}, where a strong classical pump wave with the carrier frequency $\omega_{\mathrm{p}}$ converts a signal wave with the carrier frequency $\omega_{\mathrm{s}}$ into an idler wave with the carrier frequency $\omega_{\mathrm{i}}=\omega_{\mathrm{s}}+\omega_{\mathrm{p}}$. The pump is assumed to be undepleted and its complex slowly-varying amplitude is written as $a_{\mathrm{p}}(t)=A_{\mathrm{p}}(t)\exp[i\phi_{\mathrm{p}}(t)]$, where $A_{\mathrm{p}}(t)$ and $\phi_{\mathrm{p}}(t)$ are real functions of time, being the modulus and the phase of the pump wave respectively. The signal and the idler waves are described by their slowly-varying photon annihilation operators $\hat{a}_{\mathrm{s}}(z,t)$ and $\hat{a}_{\mathrm{i}}(z,t)$, depending on the distance $z$ inside the nonlinear medium and obeying the free-field canonical commutation relations~\cite{Kolobov1999,Kolobov2007}. In the present section we neglect the group velocity mismatch in the nonlinear medium and assume that all three waves propagate inside the nonlinear crystal with the same group velocity $v_{\mathrm{g}}$. The case of different group velocities is considered in the next section.

Under these assumptions the evolution of the signal and the idler waves inside the nonlinear crystal is described by the following equations
}
\begin{eqnarray}
     \frac{\partial}{\partial z}\hat{a}_{\mathrm{s}}(z,\tau)&=
     g a_{\mathrm{p}}^{*}(\tau)\hat{a}_{\mathrm{i}}(z,\tau)
     \mathrm{e}^{-\mathrm{i}\Delta z},\label{evo a1}
     \\
     \frac{\partial}{\partial z}\hat{a}_{\mathrm{i}}(z,\tau)&=
     -g a_{\mathrm{p}}(\tau)\hat{a}_{\mathrm{s}}(z,\tau)
\mathrm{e}^{\mathrm{i}\Delta z},\label{evo a3}
\end{eqnarray}
where $g$ is the nonlinear coupling constant proportional to $\chi^{(2)}$, $\tau=t-z/v_{\mathrm{g}}$ is the retarded time, and $\Delta =k_{\mathrm{s}}+k_{\mathrm{p}}-k_{\mathrm{i}}$ is the phase mismatch between the signal, the idler, and the pump wave vectors. In order to simplify the results we consider in this section the perfect phase matching condition, $\Delta=0$. In this case Eqs.~(\ref{evo a1},\ref{evo a3}) have the following solution~\cite{Boyd},
\begin{eqnarray}
     \hat{a}_{\mathrm{s}}(L,\tau)&=
     c(\tau)\,\hat{a}_{\mathrm{s}}(0,\tau)+
     s(\tau)\mathrm{e}^{-\mathrm{i}\phi_{\mathrm{p}}(\tau)}\hat{a}_{\mathrm{i}}(0,\tau),\label{sol a1}
     \\
     \hat{a}_{\mathrm{i}}(L,\tau)&=
     -s(\tau)\mathrm{e}^{\mathrm{i}\phi_{\mathrm{p}}(\tau)}\hat{a}_{\mathrm{s}}(0,\tau)
     +c(\tau)\,\hat{a}_{\mathrm{i}}(0,\tau),\label{sol a3}
\end{eqnarray}
relating the signal and the idler annihilation operators at the output of the nonlinear crystal of length $L$, i.~e.~at $z=L$, to those at its input, $z=0$. Here we have introduced the following shorthands,
\begin{equation}
     s(\tau)=\sin[gA_{\mathrm{p}}(\tau)L],  \quad  c(\tau)=\cos[gA_{\mathrm{p}}(\tau)L].\label{coeff}
\end{equation}

{
Equations (\ref{sol a1}) and (\ref{sol a3}) describe a unitary transformation of the photon annihilation operators from the input of the nonlinear crystal to its output. It may be easily shown that these equations preserve the canonical commutation relations, which is a consequence of the relation $c(\tau)^2+s(\tau)^2=1$. Equations (\ref{sol a1}) and (\ref{sol a3}) are equivalent to the transformation performed by an ideal beam splitter with the amplitude transmission and reflection coefficients $c(\tau)$ and $s(\tau)$ respectively.
}

\subsection{The pump of the SFG-based time lens}

{
In an SFG process with an undepleted pulsed pump, the field transformation, described by Eqs.~(\ref{sol a1}) and (\ref{sol a3}), has time-dependent transmission and reflection coefficients and also a time-dependent phase factor $\exp[i\phi_{\mathrm{p}}(\tau)]$, which is determined by the phase of the pump wave. To use an SFG transformation as a time lens one needs to choose a quadratic time dependence in the phase of the pump wave: $\phi_{\mathrm{p}}(\tau)=\tau^2/2D_{\mathrm{f}}$. Such a quadratic time dependence can be produced by propagating a short pulse through a dispersive medium with the total group delay dispersion (GDD) equal to $D_{\mathrm{f}}$. In the classical  theory of temporal imaging the parameter $D_{\mathrm{f}}$ is known as the focal GDD of the time lens, and it plays a role similar to the focal distance of a conventional lens.

Now we substitute a quadratic time dependence in the phase of the pump wave in Eqs.~(\ref{sol a1}) and (\ref{sol a3}). Additionally we require that all the energy of the signal field is transmitted to the idler wavelength. This allows us to avoid a degradation of the input state nonclassicality at the frequency conversion stage. Thus, we impose the condition of a unit reflection coefficient, $s(\tau)=1$, in Eqs.~(\ref{sol a1}) and (\ref{sol a3}), which is equivalent to $gA_{\mathrm{p}}(\tau)L=\pi/2$. We suppose that these two conditions are satisfied for some interval $\tau\in [t_0,t_1]$, whose duration is known as the time lens aperture. Within this interval Eq.~(\ref{sol a3}) reads,}
\begin{equation}
     \hat{a}_{\mathrm{i}}(L,\tau)=
     -\mathrm{e}^{\mathrm{i}\tau^2/2D_{\mathrm{f}}}\hat{a}_{\mathrm{s}}(0,\tau),\label{sol a3 II}
\end{equation}
and the vacuum contribution from the photon annihilation operator $\hat{a}_{\mathrm{i}}(0,\tau)$ vanishes.

\subsection{Single-lens temporal imaging system}

{
We consider now a single-lens temporal imaging system with the magnification $M$. This system is composed of a time lens with the focal GDD $D_{\mathrm{f}}$, which is preceded by a dispersive medium with the GDD $D_{\mathrm{s}}$ for the signal wave, and is followed by another dispersive medium with the GDD $D_{\mathrm{i}}$ for the idler wave. The values of $D_{\mathrm{s}}$ and $D_{\mathrm{i}}$ are determined by the relations} $ -D_{\mathrm{i}}/D_{\mathrm{s}}=M$ and

\begin{equation}
     1/D_{\mathrm{i}}+1/D_{\mathrm{s}}=1/D_{\mathrm{f}}. \label{imaging_cond}
\end{equation}
{
The total transformation in the temporal imaging system between the input photon annihilation operator $\hat{a}_{\mathrm{s}}(\tau)$ and the output annihilation operator $\hat{a}'_{\mathrm{i}}(\tau)$ is given by a combination of propagation in the dispersive media and a nonlinear transformation determined by Eq.~(\ref{sol a3 II}), and can be written as}
\begin{equation}
     \hat{a}'_{\mathrm{i}}(\tau)=
     \frac{-1}{\sqrt{M}}\mathrm{e}^{\mathrm{i}\frac{\tau^2}{2MD_{\mathrm{f}}}} \hat{a}_{\mathrm{s}}(\tau/M). \label{imaging_eq}
\end{equation}
{
Equation~(\ref{imaging_eq}) describes a magnification of the quantum input signal $\hat{a}_{\mathrm{s}}(\tau)$ by a factor of $M$ and is valid within some interval $\tau\in [t_0',t_1']$, whose duration may be much larger than the time lens aperture in the high magnification limit.
}

\subsection{Temporal imaging system for squeezed light}

{
In this subsection we consider a temporally broadband squeezed state of light as an input state of the temporal imaging system. Such light can be generated by a traveling-wave optical parametric amplifier (OPA) in a second-order nonlinear crystal~\cite{Kolobov1999} and is described in terms of Fourier amplitudes of the field operators,
}
\begin{equation}
     \hat{a}_{\mathrm{s}}(z,\Omega)=\int \hat{a}_{\mathrm{s}}(z,t)\mathrm{e}^{\mathrm{i}\Omega t}\mathrm{dt}. \label{Fourier}
\end{equation}
{
The transformation of the field operators in an OPA is given by the following equation,
}
\begin{equation}
     \hat{a}_{\mathrm{s}}(l,\Omega)=
     U(\Omega)\hat{a}_{\mathrm{s}}(0,\Omega)+V(\Omega)\hat{a}^{\dag}_{\mathrm{s}}(0,-\Omega), \label{Bogolubov}
\end{equation}
{
where $l$ is the length of the OPA crystal, and $U(\Omega)$ and $V(\Omega)$ are the complex coefficients depending on the parametric gain of the OPA and its phase-matching conditions \cite{Kolobov1999}.

Transformation Eq.~(\ref{Bogolubov}) produces a broadband squeezed vacuum at the output of the OPA from the broadband vacuum input state. The effect of quadrature squeezing is observed by means of the balanced homodyne photodetection, where the measured state is mixed with a strong monochromatic local oscillator (LO). The homodyne photocurrent noise spectrum, normalized to the shot-noise level $S(\Omega)=(\delta i)^2_{\mathrm{\Omega}}/\langle i\rangle$, is known as the squeezing spectrum. Considering for simplicity the unit photodetection efficiency, we can write the squeezing spectrum as follows,
}
\begin{equation}
     S(\Omega)=1+\int\langle: \hat{X}_\varphi(t)\hat{X}_\varphi(t+\tau): \rangle \mathrm{e}^{i\Omega\tau}d\tau, \label{squeezingdef}
\end{equation}
where the colon stands for normal ordering and $\hat{X}_\varphi(t) = \hat{a}_{\mathrm{s}}(t)\mathrm{e}^{-i\varphi} +\hat{a}_{\mathrm{s}}^\dagger(t)\mathrm{e}^{i\varphi}$ is the operator of the field quadrature corresponding to the phase of LO, which we have denoted as $\varphi$.

Substituting Eq.~(\ref{Bogolubov}) into Eq.~(\ref{squeezingdef}) we obtain the spectrum of squeezing of the OPA radiation:
\begin{equation}
     S_{\mathrm{s}}(\Omega)=\cos^2\theta(\Omega)\mathrm{e}^{2r(\Omega)} +\sin^2\theta(\Omega)\mathrm{e}^{-2r(\Omega)}, \label{squeezing}
\end{equation}
where $\theta(\Omega)=\psi(\Omega)-\varphi$, with the squeezing angle $\psi(\Omega)$ and the squeezing parameter $r(\Omega)$ being given by,
\begin{eqnarray}
     &\psi(\Omega)=
     \frac{1}{2}\arg[V(\Omega)/U(\Omega)],\label{sq_angle}
     \\
     &\exp[\pm r(\Omega)]=|U(\Omega)|\pm|V(\Omega)|.\label{sq_param}
\end{eqnarray}

{
Now we consider the transformation of the squeezed field in a temporal imaging system.} To calculate the spectrum of squeezing we assume first, that the pump intensity is constant during its interaction with the signal, $A_{\mathrm{p}}(t)=A_{\mathrm{p0}}$, and second that the phase modulation of the field at the output of the temporal imaging system is negligible or compensated. Under these assumptions,
{
substituting the Fourier transform of the state Eq.~(\ref{Bogolubov}) into Eq.~(\ref{imaging_eq}), we obtain the following squeezing spectrum for the operator $\hat{a}'_{\mathrm{i}}(\tau)$},
\begin{equation}
     S_{\mathrm{i}}(\Omega)=1-\eta+\eta[\cos^2\theta(\tilde\Omega)\mathrm{e}^{2r(\tilde\Omega)}+
     \sin^2\theta(\tilde\Omega)\mathrm{e}^{-2r(\tilde\Omega)}], \label{squeezing1}
\end{equation}
with $\tilde\Omega = |M|\Omega$ and $\eta=\sin^2(gA_{\mathrm{p0}}L)$ being the efficiency factor of the time lens equal to the intensity reflection coefficient in Eqs.~(\ref{sol a1},\ref{sol a3}).

\begin{figure}[t!]
\centering
\includegraphics[width=\columnwidth]{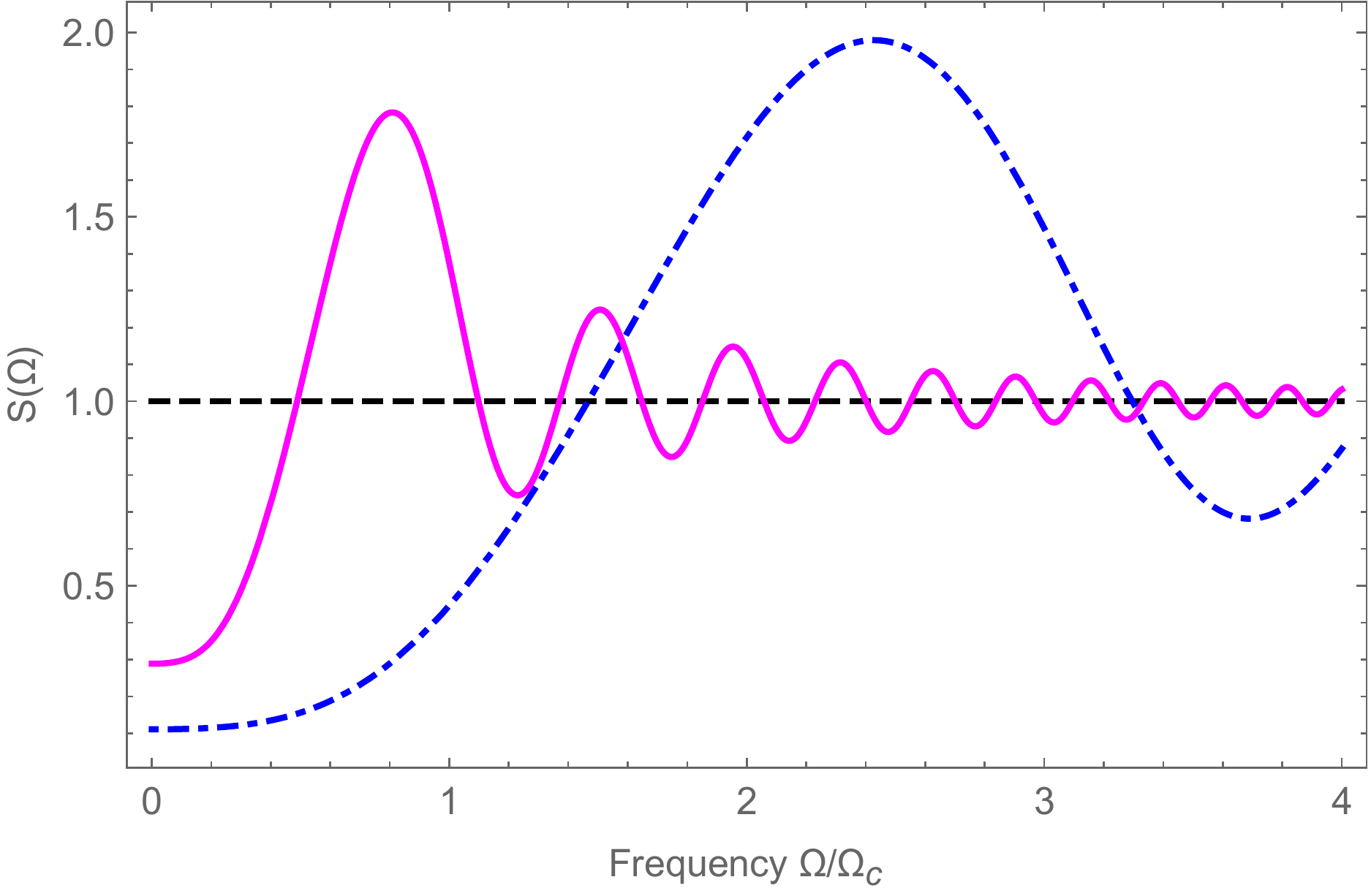}
\caption{Squeezing spectrum of broadband squeezed light at the output of the OPA (dash-dotted) and after a temporal imaging system with the magnification $M=-3$ and the efficiency factor $\eta=0.8$ (solid). For both curves $\exp[ r(0)] = 3$.}
\label{fig:Spectrum}
\end{figure}

In Fig.~\ref{fig:Spectrum} we show this squeezing spectrum for $\exp[r(0)]=3$, where $r(0)$ is the maximum value of the squeezing parameter at $\Omega=0$. One can see from Fig.~\ref{fig:Spectrum} that at low frequencies, $\Omega<\Omega_{\mathrm{c}}$, the quantum fluctuations of the photocurrent are reduced below the shot-noise level. This characteristic frequency is defined as $\Omega_{\mathrm{c}}=(\beta^{(c)}_{\mathrm{2}}l)^{-1/2}$, where $\beta^{(c)}_{\mathrm{2}}$ is the GVD coefficient of the OPA. Fig.~\ref{fig:Spectrum} illustrates two main effects of a temporal imaging system on the squeezed light, described by Eq.~(\ref{squeezing1}). First, the squeezing spectrum at the output of this system is an $|M|$ times compressed version of that at the input. Second, for $\eta=0.8$ (and generally for $\eta<1$), the vacuum fluctuations entering into the temporal lens from the open port, deteriorate the squeezing at its output.

In summary for this section, we have shown that for operating in a quantum regime, an SFG-based time lens should have a quadratic in time dependence of the pump phase and provide an almost unit conversion efficiency from the signal to the idler wavelength.

\section{Temporal imaging system in the quantum regime}

\subsection{General description of the temporal imaging system}
In this section we analyze a temporal imaging system on the basis of an SFG-based time lens described in the previous section. As has been mentioned in the Introduction, a temporal imaging system is aimed at stretching (or compressing) an optical waveform. We define this waveform, which we call also ``object field'', as temporal variation of light between the times $t_0$ and $t_0+T_{F}$, where $t_0$ is some starting time and  $T_{F}$ is the temporal duration of the range to be imaged, the field of view (FOV) \cite{Bennett2000a,Bennett2000b}. The temporal imaging system includes as a key element an SFG time lens, characterized by the focal GDD $D_{\mathrm{f}}$ and the aperture $T$. The system also includes a dispersive area with GDD $D_{\mathrm{s}}$ to be passed by the signal before the lens, and a dispersive area with GDD $D_{\mathrm{i}}$ to be passed by the idler field after the lens. For imaging with a magnification $M=-D_{\mathrm{i}}/D_{\mathrm{s}}$, it is necessary that the dispersive elements before and after the temporal lens satisfy Eq.~(\ref{imaging_cond}).

It is easy to find that a high magnification, $|M|\gg 1$, requires $D_{\mathrm{s}}\approx D_{\mathrm{f}}$, which in spatial optics corresponds to placing the object close to the focal plane.

In the next subsections we discuss the important limitations imposed on the experimental conditions for realizing such a temporal imaging system in the quantum regime.

\subsection{Preparation of the pump}
Let us suppose that the bandwidth of the signal field is limited to $\delta\omega_{\mathrm{s}}$. This is, for example, the case when a signal is imparted on the continuous-wave squeezed light by means of a modulator with that bandwidth, or when the input field is emitted by a physical system with a characteristic time $\tau_0=2\pi/\delta\omega_{\mathrm{s}}$. For the sake of simplicity we can model the input field as composed of $N=T_{F}/\tau_0$ temporal pixels, each pixel being a Gaussian pulse of width $\tau_0$, the distance between the centers of the adjacent pixels being also $\tau_0$. Here and below we understand the width of a Gaussian spectrum as the full width at half-maximum (FWHM) for the amplitude, and the temporal duration of a Gaussian pulse as the FWHM of its amplitude, multiplied by $2\pi/(8\ln 2)\approx 1.13$. We accept that the carrier frequency, denoted as $\omega_{\mathrm{s}}$, is the same for all pixels. This condition corresponds in the spatial imaging to the requirement that the object is illuminated by a plane wavefront.

After passing through the input dispersive element, each pixel is stretched to the duration $\Delta t_0=D_{\mathrm{s}}\delta\omega_{\mathrm{s}}$, which we accept to satisfy the relation $\Delta t_0\gg \tau_0$, typical for temporal imaging systems. The nearby pixels will be highly overlapping but if the carrier frequency $\omega_{\mathrm{s}}$ is the same for all pixels, the distance between their peaks will be $\tau_0$, as initially. Thus, the whole input field is stretched to the duration $T_{\mathrm{s}} = \Delta t_0+(N-1)\tau_0$. This time range should be accommodated by the temporal lens aperture $T$, which we accept to be equal to the duration of the stretched waveform, $T=T_{\mathrm{s}}$.

\begin{figure}[t!]
\centering
\includegraphics[width=\columnwidth]{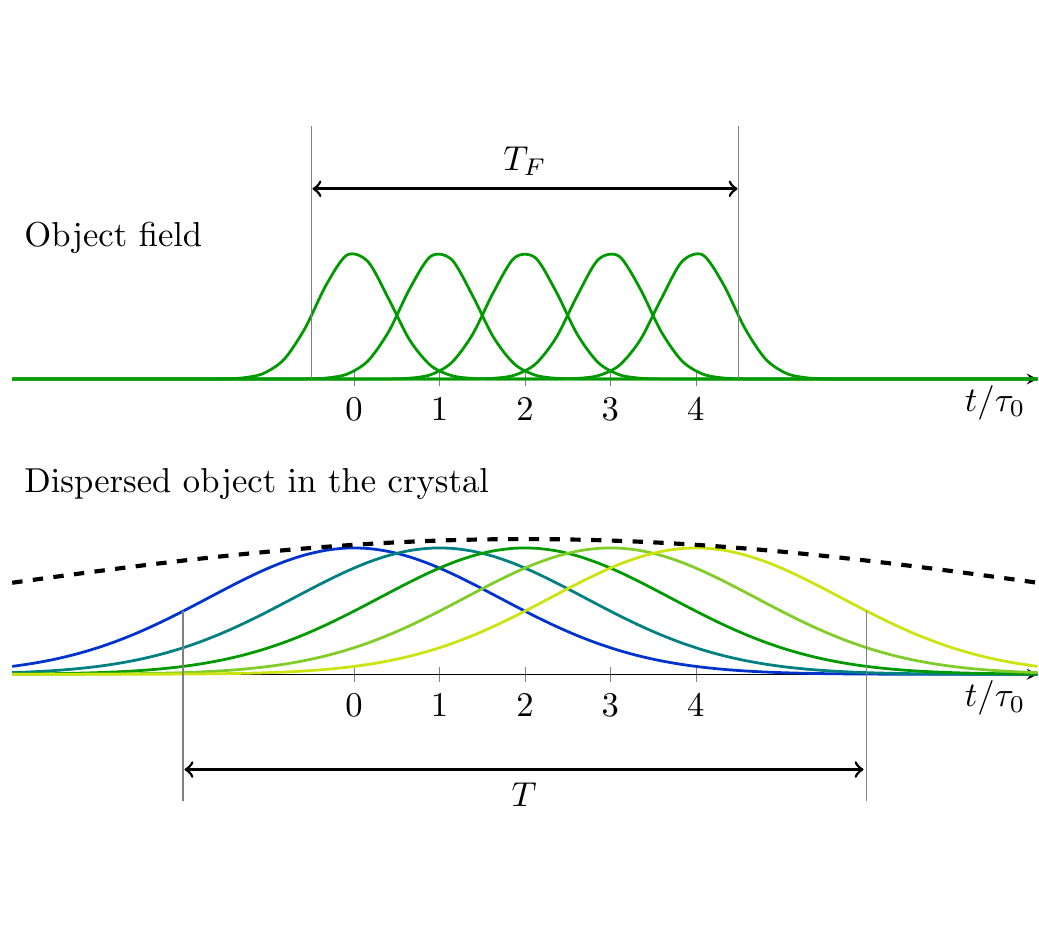}
\caption{Representation of the field in a temporal imaging system as as a collection of temporal pixels: (above) field envelopes of the object, (below) field envelopes inside the nonlinear crystal, after being stretched in a dispersive element. Four-time stretching is shown as illustration, practical values reach several orders of magnitude. Different colours of the pixels inside the crystal show different frequency shifts received from the interaction with the pump. Black dashed line is the pump envelope.}
\label{fig:Pulses}
\end{figure}

The quantum regime of the SFG time lens requires that the modulus of the pump satisfies the condition at which the conversion efficiency from the signal to the idler wavelength is close to unity. As consequence, we arrive at the requirement that the pump modulus should be almost uniform within the aperture $T$ (see Fig.~\ref{fig:Pulses}). If this requirement is not met, only the central pixel is imaged in the lossless regime. For all other pixels the operation of the lens is lossy, and therefore, the output intensity is lower than that of the central pixel, and also the degree of squeezing deteriorates dramatically.

The pump for a time lens is typically prepared by passing a pulse of duration $\tau_p$ (bandwidth $\delta\omega_{\mathrm{p}}=2\pi/\tau_p$) through a dispersive element with a GDD equal to $-D_{\mathrm{f}}$, where it acquires the necessary quadratic phase dependence. In this case the pump pulse at the entrance of the nonlinear crystal has the duration $\Delta t_p = D_{\mathrm{f}}\delta\omega_{\mathrm{p}}$, and the requirement of an almost uniform pump modulus becomes $\Delta t_p \gg T$, or $\delta\omega_{\mathrm{p}} \gg \delta\omega_{\mathrm{pa}}$, where
\begin{equation}\label{active}
\delta\omega_{\mathrm{pa}} = \delta\omega_{\mathrm{s}}\left(1+\frac1{|M|}\right)+(N-1)\frac{\tau_0}{D_{\mathrm{f}}}.
\end{equation}
is the active pump bandwidth, the part of the pump bandwidth, which participates in the nonlinear interaction with the signal. The rest of the pump is either delayed (lower frequencies) or advanced (higher frequencies) and does not overlap with the signal pulse in the crystal.

Alternatively, a pump pulse with the bandwidth, given by Eq.~(\ref{active}) can be prepared by means of a pulse shaper. In this case the pump pulse should be shaped to the temporal dependence

\begin{equation}\label{Ap}
A_p(t)=A_{p0}e^{it^2/(2D_{\mathrm{f}})}\Pi(t/T)
\end{equation}
where $A_{p0}$ is a constant and $\Pi(t)$ is a rectangular function. This method of preparation involves a more complicated active technique, but has two advantages: it requires longer initial pulses and provides a pump with sharp edges, necessary for the output field being the image of the object field, not mixed with the light before or after the field of view.

\subsection{Phase-matching condition}

The phase mismatch for three waves in the SFG crystal is given by
\begin{equation}\label{Dk}
\Delta(\Omega_s,\Omega_i)=k_s(\Omega_s)+k_p(\Omega_i-\Omega_s)-k_i(\Omega_i),
\end{equation}
where $\Omega_s$, $\Omega_p$, and $\Omega_i$ are detunings from the central frequencies for the signal, the pump and the idler waves, while $k_s(\Omega)$, $k_p(\Omega)$, and $k_i(\Omega)$ are the wavevectors of these waves at the detuning $\Omega$. We accept that the angles between the three propagating waves are small enough to be disregarded and that the phase-matching at the central frequencies is perfect, $\Delta(0,0)=0$. For small enough signal and idler bandwidths we can decompose the wavevectors as functions of frequency in the Taylor series and leave just the linear terms:
\begin{equation}\label{Dk2}
\Delta(\Omega_s,\Omega_i)\approx (k_s'-k_p')\Omega_s+(k_p'-k_i')\Omega_i,
\end{equation}
where $k_m'$ denotes the first derivative at $\Omega=0$ for the function $k_m(\Omega)$, $m=p,s,i$.

The first term in the r.h.s. of Eq.~(\ref{Dk2}) corresponds to the relative group delay of the pump and the signal waves, the second one -- to that of the pump and the idler. For a lossless operation of the time lens it is necessary that both these terms are close to zero.

The term $k_s'-k_p'$ can be nullified if one chooses for the time lens a frequency degenerate type-I SFG, where the signal and the pump waves are ordinary, while the idler wave is extraordinary \cite{Bennett1999}. In this case the signal and the pump waves travel in the crystal at the same group velocity, while the idler wave is delayed with respect to them. The total delay time is
\begin{equation}\label{taud}
\tau_i=|k_p'-k_i'|L,
\end{equation}
where $L$ is the crystal length. An almost perfect phase-matching requires this delay to be small compared to the inverse of the idler bandwidth, which we denote $\delta\omega_{\mathrm{i}}$:
\begin{equation}\label{limiti}
\tau_i \ll \frac{2\pi}{\delta\omega_{\mathrm{i}}}.
\end{equation}

The limitation on the signal bandwidth is much less stringent and can be obtained when decomposing the phase mismatch function up to the second order in $\Omega_s$:
\begin{equation}\label{Dk4}
\Delta(\Omega_s,\Omega_i)\approx k_s''\Omega_s^2+(k_p'-k_i')\Omega_i,
\end{equation}
where $k_s''$ is the second derivative at zero for the function $k_s(\Omega)$, which is the same for the signal and the pump. The quadratic in frequency term describes an additional spreading of the pixel in the crystal with a total GDD $k_s''L$. The spreading time can be estimated as
\begin{equation}\label{taus}
\tau_s = \sqrt{k_s''L}.
\end{equation}
The spreading can be disregarded if $\tau_s \ll \tau_0$, wherefrom we obtain the limitation
\begin{equation}\label{limits}
\tau_s \ll \frac{2\pi}{\delta\omega_{\mathrm{s}}}.
\end{equation}

If the conditions Eq.~(\ref{limiti}) and (\ref{limits}) are not satisfied, the output of the imaging system is a filtered image of the input. A quantum image is much more sensitive to the filtering than a classical one, because of the vacuum added to the output field. This vacuum can destroy or significantly deteriorate the quantum features of the imaged field.

\subsection{The output field}

Different pixels arrive at the nonlinear crystal at different times and see different slopes of the pump phase variation. Therefore, during the SFG process they receive different central frequency shifts. The temporal distance between the adjacent pixels is $\tau_0$, therefore the difference of central frequencies is $\Delta\omega=\ddot\phi_{\mathrm{p}}\tau_0=\tau_0/D_{\mathrm{f}}$. After leaving the nonlinear crystal the idler beam is a sequence of $N$ stretched pixels of spectral width $\delta\omega_{\mathrm{s}}/|M|$ each with the spectral distance between the adjacent pixels $\Delta\omega$. The total spectral width of the idler beam is thus
\begin{equation}\label{wi}
\delta\omega_{\mathrm{i}} = \frac{\delta\omega_{\mathrm{s}}}{|M|}+(N-1)\frac{\tau_0}{D_{\mathrm{f}}}.
\end{equation}
Comparing Eq.~(\ref{wi}) with Eq.~(\ref{active}) we arrive at the fundamental relation of lossless temporal imaging:
\begin{equation}\label{rel}
\delta\omega_{\mathrm{pa}} = \delta\omega_{\mathrm{s}}+\delta\omega_{\mathrm{i}}.
\end{equation}

\begin{figure}[t!]
\centering
\includegraphics[width=\columnwidth]{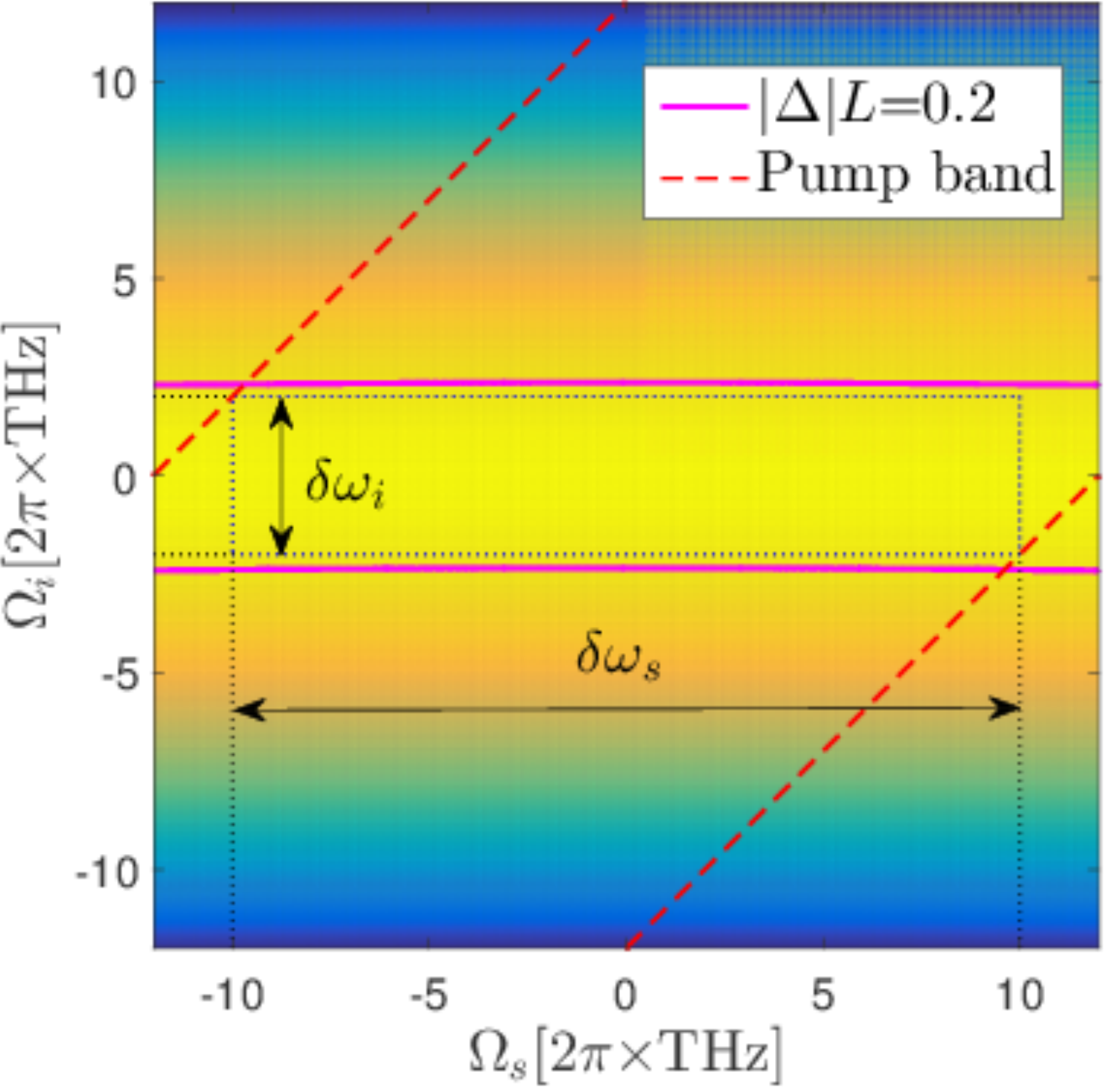}
\caption{Phase mismatch of three waves in the SFG crystal (color map) with the area where it is negligible (between the solid lines) versus pump function (dashed lines) in the space of
signal and idler detunings from the central frequencies. The crystal is a 500 $\mu$m thick beta-barium borate (BBO) crystal, cut at 28.1$^{\circ}$ for a non-collinear type-I phase-matching of pump and signal waves at 830 nm and the idler wave at 415 nm, as in Ref.~\cite{Bennett2001}. Dotted lines show the signal and the idler bands. }
\label{fig:PM+pump}
\end{figure}

Equation (\ref{rel}) can be understood from the diagram of the phase mismatch in the SFG crystal, shown in Fig.~\ref{fig:PM+pump}. The signal and the idler bands (dotted lines) are chosen such that the area of their intersection lies completely inside the area of negligible mismatch (between the solid lines). The signal and idler waves with the detunings $\Omega_s$ and $\Omega_i$ interact with the pump of detuning $\Omega_p=\Omega_i-\Omega_s$. The dashed lines show the spectral area where the pump modulus should be constant. The distance between the dashed lines is equal to $\delta\omega_{\mathrm{pa}}/\sqrt{2}$ and Eq.~(\ref{rel}) can be deduced easily from planimetric considerations.

The pixels at the output of the nonlinear crystal are negatively chirped, and afterwards they are compressed during their passage through the dispersive medium with the GDD $D_{\mathrm{i}}$. At the output of this medium each pixel has the transform-limited duration $|M|\tau_0$, and the distance between the adjacent pixels is $ D_{\mathrm{i}}\Delta\omega=|M|\tau_0$. Thus, the output pixels represent a stretched (and inverted) copy of the input pixels with the additional linear shift of the carrier frequencies. If the operation of the imaging system is lossless, then no vacuum is admixed to the output field and the quantum structure is not degraded.

\subsection{Quantum field of view}

Equations~(\ref{active}), (\ref{limiti}), (\ref{limits}) and (\ref{wi}) determine the resolution and the FOV of a given temporal imaging system. For a given nonlinear crystal, the resolution is limited by Eq.~(\ref{limits}), but also by Eqs.~(\ref{limiti},\ref{wi}), because even for imaging just one pixel, the signal bandwidth cannot surpass that of the idler more than $|M|$ times. The number of pixels in the FOV is determined from Eq.~(\ref{wi}) as
\begin{equation}\label{N}
N = 1 + \frac{D_{\mathrm{f}}}{\tau_0} \left(\delta\omega_{\mathrm{i}}-\frac{\delta\omega_{\mathrm{s}}}{|M|}\right).
\end{equation}
For sufficiently high magnification, when $|M|\gg \tau_i/\tau_0$ the second term in parenthesis can be disregarded. In this limit and in the practically important case of high number of pixels ($N\gg1$), the field of view is limited by
\begin{equation}\label{TF}
T_F^{(\mathrm{q})} \ll \frac{2\pi D_{\mathrm{f}}}{\tau_i},\quad T_F^{(\mathrm{q})} \ll \frac{2\pi D_{\mathrm{f}}}{\tau_p},
\end{equation}
where the second limitation follows from the requirement of a uniform pump modulus in the lens aperture and the superscript $(\mathrm{q})$ stands for quantum regime. We see that the main limiting factor for the field of view is the group velocity dispersion in the nonlinear crystal. We note also that the definition of the FOV in quantum temporal imaging is different of that for its classical counterpart \cite{Bennett2000b}. The classical FOV is defined as a time range in the input waveform, where pixels are degraded by the imaging system less than two times. In a typical configuration for classical temporal imaging it is the width of the product of the pump envelope and the filtering function, i.e.
\begin{equation}\label{TFc}
T_{F}^{(\mathrm{cl})} \approx \frac{2\pi D_{\mathrm{f}}}{\sqrt{\tau_p^2+\tau_i^2}}.
\end{equation}
Comparing Eq.~(\ref{TFc}) to Eq.~(\ref{TF}) we arrive at the conclusion that the quantum FOV is much smaller than its classical counterpart for the same temporal imaging system, $T_{F}^{(\mathrm{q})}\ll T_{F}^{(\mathrm{cl})}$. That is not surprising, because the quantum FOV is the temporal range where the degradation of pixels is negligible.

\section{Conclusions}
We have considered application of an SFG-based temporal imaging system to an optical temporal waveform which contains light in a squeezed state. We have analyzed the restrictions imposed on the pump and the phase matching condition in the crystal in order to preserve squeezing in the output signal. We have shown that modification of the notion of the FOV in the quantum case is necessary, and that the quantum FOV is much narrower than the classical one for the same temporal imaging system. Our results are of crucial importance for temporal stretching and compressing of squeezed optical pulses, used in quantum-enhanced metrology and quantum communications.

\section*{Acknowledgement}
This work was supported by the European Union's Horizon 2020 research and innovation programme under grant agreement No 665148 (QCUMbER).

\end{document}